\documentclass[aip,twocolumn,showpacs,superscriptaddress,preprintnumbers,amsmath,amssymb]{revtex4}
\usepackage{amsmath}
\usepackage{graphicx}
\usepackage{dcolumn}
\usepackage{bm}
\usepackage{float}

\begin{document}

\title{Simulation of relativistically colliding laser-generated electron flows}


\author{X. H. Yang}
\affiliation{Centre for Plasma Physics, School of Mathematics and Physics,
Queen's University of Belfast, Belfast BT7 1NN, United Kingdom}
\affiliation{College of Science, National University of Defense Technology,
Changsha 410073, China}

\author{M. E. Dieckmann}\thanks{Electronic mail: Mark.E.Dieckmann@itn.liu.se}
\affiliation{Centre for Plasma Physics, School of Mathematics and Physics,
Queen's University of Belfast, Belfast BT7 1NN, United Kingdom}

\author{G. Sarri}\affiliation{Centre for Plasma Physics, School of
Mathematics and Physics,  Queen's University of Belfast, Belfast BT7 1NN,
United Kingdom}

\author{M. Borghesi}
\affiliation{Centre for Plasma Physics, School of Mathematics and Physics,
Queen's University of Belfast, Belfast BT7 1NN, United Kingdom}
\affiliation{Institute of Physics of the ASCR, ELI-Beamlines project, Na Slovance 2, 18221 Prague, Czech Republic}


\date{\today}

\begin{abstract}
The plasma dynamics resulting from the simultaneous impact, of two equal, ultra-intense laser pulses, in two spatially separated spots, onto  a dense target  is studied via particle-in-cell (PIC) simulations. The simulations show that electrons accelerated to relativistic speeds, cross the target and exit at its rear surface. Most energetic electrons are bound to the rear surface by the ambipolar electric field and expand along it. Their current is closed by a return current in the target, and this current configuration generates strong surface magnetic fields.
The two electron sheaths collide at the midplane between the laser
impact points. The magnetic repulsion between the counter-streaming electron beams separates them
along the surface normal direction, before they can thermalize through other beam instabilities. This
magnetic repulsion is also the driving mechanism for the beam-Weibel (filamentation) instability,
which is thought to be responsible for magnetic field growth close to the internal shocks of
gamma-ray burst (GRB) jets. The relative strength of this repulsion compared to the competing
electrostatic interactions, which is evidenced by the simulations, suggests that the filamentation
instability can be examined in an experimental setting.
\end{abstract}

\pacs{52.38Kd, 52.38.Fz, 52.65.Rr}

\maketitle

\section{\label{sec:level1}INTRODUCTION\protect}

The dynamics of the transient electromagnetic field driven by relativistic electronic surface
currents, which are generated by the interaction of ultra-intense laser beams with solids
\cite{s1,s3,s4}, are of relevance to important applications such as laser-driven ion acceleration \cite{s5,s5a} and fast ignition
\cite{s6}. The surface magnetic and electrostatic fields can be of the order of $\sim10^3$T and
$\sim 10^{12}$V/m \cite{s6a,s7,s8}, respectively, during the ultra-intense ($I>10^{19}$W/cm$^2$)
laser-solid interaction, and such fields can last several tens of picoseconds after the laser pulse
ends. Charging of a laser-irradiated target, which is attributed to the escape of hot electrons generated during
the laser-plasma interaction, has been detected in several experiments \cite{s9,s9a} using proton probing techniques.
Because of their confinement by the sheath electric field and surface magnetic field, most of the
hot electrons are bound to the target surface \cite{s4,s10,s11,s12} as they move away from the
interaction region. The surface current is closed by the return current within the target and this
current loop encloses a strong magnetic field. This magnetic field generation mechanism is known
as the fountain effect. Recent experiments \cite{s13,s14} have demonstrated that, during the ultra-intense laser irradiating the target, the magnetic
field tied to the electron current expands along the solid's surface at a speed close to that of light.

If two ultra-intense laser pulses of comparable intensity irradiate the target's surface at two separate
points simultaneously, they produce two relativistic electron flows at the rear surface. Both flows collide
head-on half-way between the two interaction points at the planar rear surface. The superposition of
their respective currents, which are oppositely directed and comparable in magnitude, initially
results in a net current density, which is low compared to the current density of each electron beam. This
reduction of the current density results in a weakened magnetic field in the beam overlap region. The
time-evolution of the counter-streaming beams and of the electric and magnetic fields, which result in
their thermalization, depends on the nature of the instabilities and processes that are involved.

Previous related experiments have examined the simultaneous interaction of two
or more laser pulses with a solid target. The laser pulses had intensities of
$10^{14}-10^{15}$W/cm$^{2}$ \cite{s14a,s14b}. In these experiments, the expansion of plasma
bubbles at the front surface of the target was examined with the aim to study magnetic
reconnection of the megagauss-fields, which encircle the individual plasmas
and confine them.
The moderate laser intensity has limited the particle flow speeds to the nonrelativistic regime.
Such flows tend to thermalize electrostatically.

Here the speeds of the electron sheaths, which form before ion jets are
launched, are relativistic and the magnetic fields sustained by their
currents are confined to a narrow layer close to the surface. The
instabilities that result in the thermalisation of the relativistic
fast electron flows differ from the nonrelativistic ones in Ref.
\cite{s14a,s14b} that involve electrons and ions.

The counter-streaming relativistic electrons give rise to a wide range
of beam instabilities in this overlap layer. The potentially large number
of instabilities can be reduced to three, if we assume that only the
electrons interact, that the thermal spread of the electron speeds is
smaller than the difference of the mean speeds of both beams and that
magnetic field effects in the beam overlap region are negligible.
Negligible here means that the electron gyrofrequency is small compared to the
growth rates of the instabilities. We obtain under these conditions the
two-stream instability, the oblique mode instability and the filamentation
or beam-Weibel instability. These instabilities are discussed in detail in
Ref. \cite{New1}. Here we summarize their properties.

Let $\mathbf{k}$ and $\mathbf{v}_b$ be the wave vector of the growing waves
and the difference vector between the mean velocities of the counter-streaming
electron beams, respectively. Let $\alpha$ be the angle between $\mathbf{k}$
and $\mathbf{v}_b$. The two-stream instability \cite{Thode} is purely electrostatic
for $\mathbf{k} \parallel \mathbf{v}_b$ ($\alpha = 0$). It saturates by the formation
of electron phase space holes in the nonrelativistic \cite{Roberts} and relativistic
regimes \cite{New1}. The coalescence instability \cite{Roberts} triggers the collapse
of these nonlinear structures, which thermalizes the electron flow. What remains from
the two counter-streaming electron beams is a single hot electron distribution with a
low mean speed modulus. The net current is small and the magnetic field it drives is
weak. The two-stream instability grows slower than the oblique mode instability if
$|\mathbf{v}_b|\approx c$. The oblique modes have a mixed polarity and they are
almost electrostatic. Their wave vectors $\mathbf{k}$ form angles $0 < \alpha <
90^\circ$ with respect to $\mathbf{v}_b$ and they become two-stream unstable
modes if $\alpha =0$. They also saturate by electron trapping \cite{Oblique}. Both
instabilities thermalize the electrons electrostatically.

If the counter-streaming electron beams have a similar density and temperature and if
they collide at a moderately relativistic speed, then both electrostatic instabilities are
outgrown by the filamentation instability \cite{Filament}. The latter is driven by the
magnetic repulsion of electrons, which move into opposite directions. It saturates by
magnetic trapping \cite{Davidson} and its final state are current channels, which contain
charged particles that have the same current direction. These channels have a diameter
of the order of the electron skin depth and are surrounded by strong magnetic fields. The
plasma flow is eventually thermalized through the repeated mergers of the current channels.
However, the magnetic fields remain strong in a broad spatial interval
\cite{New3,Silva,Nishikawa,Murphy}.

Our 2D PIC simulation, which uses the 2D3V code LAPINE \cite{s15}, resolves a
cross-section of the target that contains the surface normal. The electron
flow at the rear surface is thus confined to a layer close to the one-dimensional
surface with a width that is comparable to an electron skin depth. The colliding
electron sheaths can drive two instabilities in this geometry. The two-stream
instability, which results in electrostatic oscillations along the surface direction,
and the filamentation instability, which separates both electron beams along the
surface normal. Our simulation demonstrates that for the selected initial conditions
the magnetic interaction is more important than the electrostatic one. A laser-plasma
experiment with similar initial conditions would thus allow us to drive the filamentation
instability on the two-dimensional surface of the target. The experimentally measured
data can be compared to the existing 2D and 3D simulation studies (See for example
 \cite{New3,Silva,Nishikawa,Murphy}) of colliding leptonic flows and test their validity. Studying
this instability with a controlled laboratory experiment would provide us with a much needed
better understanding of the magnetic field generation within gamma-ray burst (GRB) jets,
which are spectacular releases of energetic electromagnetic radiation at cosmological
distances \cite{New2}. A direct comparison of the filaments observed in a controlled
laboratory experiment and those believed to exist close to the internal shocks of GRBs is
possible if the plasma is collisionless, because then all observables can be scaled between
both systems with the help of the electron plasma frequency \cite{Honda}.

This paper is structured as follows. Section 2 discusses the numerical scheme
and the initial conditions. The results are presented in Section 3 and they
are discussed in Section 4.

\section{SIMULATION MODEL}
Particle-in-cell simulation codes solve the numerical approximation of
Maxwell's equations on a grid and they approximate the plasma by an ensemble
of computational particles. The particle momenta are updated with the
help of the relativistic Lorentz force equation. This simulation method is
discussed in detail in Ref. \cite{Dawson}. We restrict our simulation to a
2D geometry.

\begin{figure}\suppressfloats
\includegraphics[width=6cm]{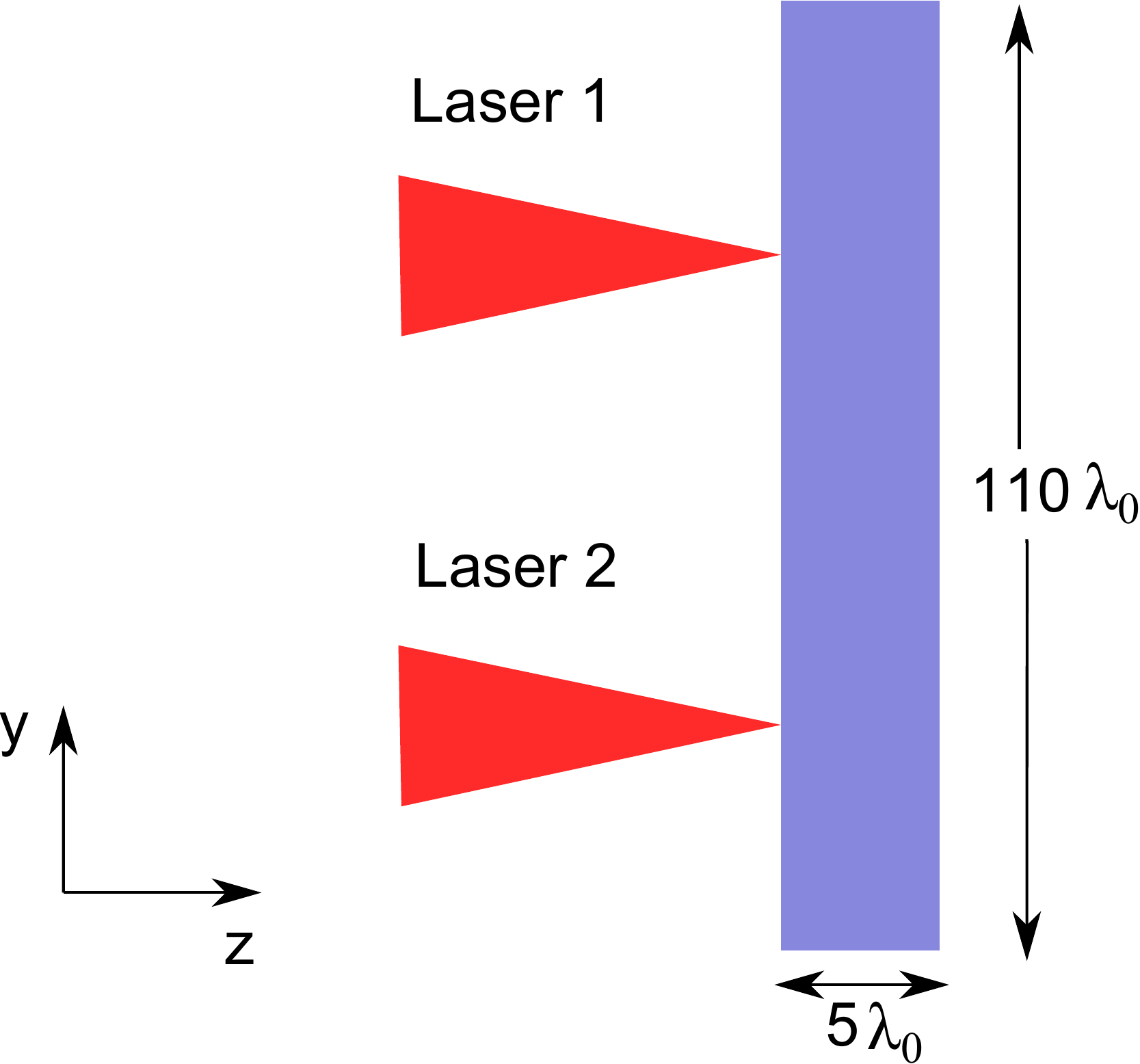}%
\caption{\label{f1} (Color online) The simulation setup: The target (blue)
has a thickness of 5$\lambda_0$ along the z-direction and is located between
$z=10 \lambda_0$ and $15 \lambda_0$ in the simulation box. The target is hit
by two equal laser pulses (red).}
\end{figure}

Our initial and boundary conditions are as follows: Two equal laser pulses
hit a thick plasma slab simultaneously and at normal incidence as depicted
in Fig. \ref{f1}.
The target with a width of 110$\lambda_0$ and thickness of 5$\lambda_0$
is initially located between  $z=10\lambda_0$ and $15\lambda_0$, where
$\lambda_0=1\mu$m is the laser wavelength. The target consists of an
initially charge-neutral mixture of electrons and Al$^{10+}$ with mass
$m_i=27m_p$, where $m_p=1836m_e$ is the proton mass. The initial density
of the target is set to 50$n_c$, where $n_c=1.12\times10^{21}$cm$^{-3}$ is
the critical density. The initial temperatures of the electrons and ions
are both set to $1$ keV for computational reasons. The simulation box
size $L_z \times L_y = 30\lambda_0\times110 \lambda_0$ is resolved by
$1500 \times 5500$ cells, and the electrons and Aluminium ions are
represented by 64 and 16 computational particles per cell, respectively. Two
$p$-polarized laser pulses are incident normally from the left boundary
and focused on the target at $y=\pm30\lambda_0$, respectively. The midpoint
between the laser pulses defines $y=0$. The pulses rise up in the first
$4T_0$ with a Gaussian profile and then maintain the peak intensity for
$40T_0$, where $T_0\sim 3.3$ fs is the laser cycle. The maximum intensity
of the pulses is $5\times10^{19}$W/cm$^2$, corresponding to $a_0=6$. The
laser pulses have a Gaussian spatial profile with a spot radius of
$5\lambda_0$. The time step is $0.007T_0$. For both the transverse and
longitudinal boundaries, absorbing boundary conditions are used for the
fields and particles.

\section{RESULTS}
\begin{figure}\suppressfloats
\includegraphics[width=8.7cm]{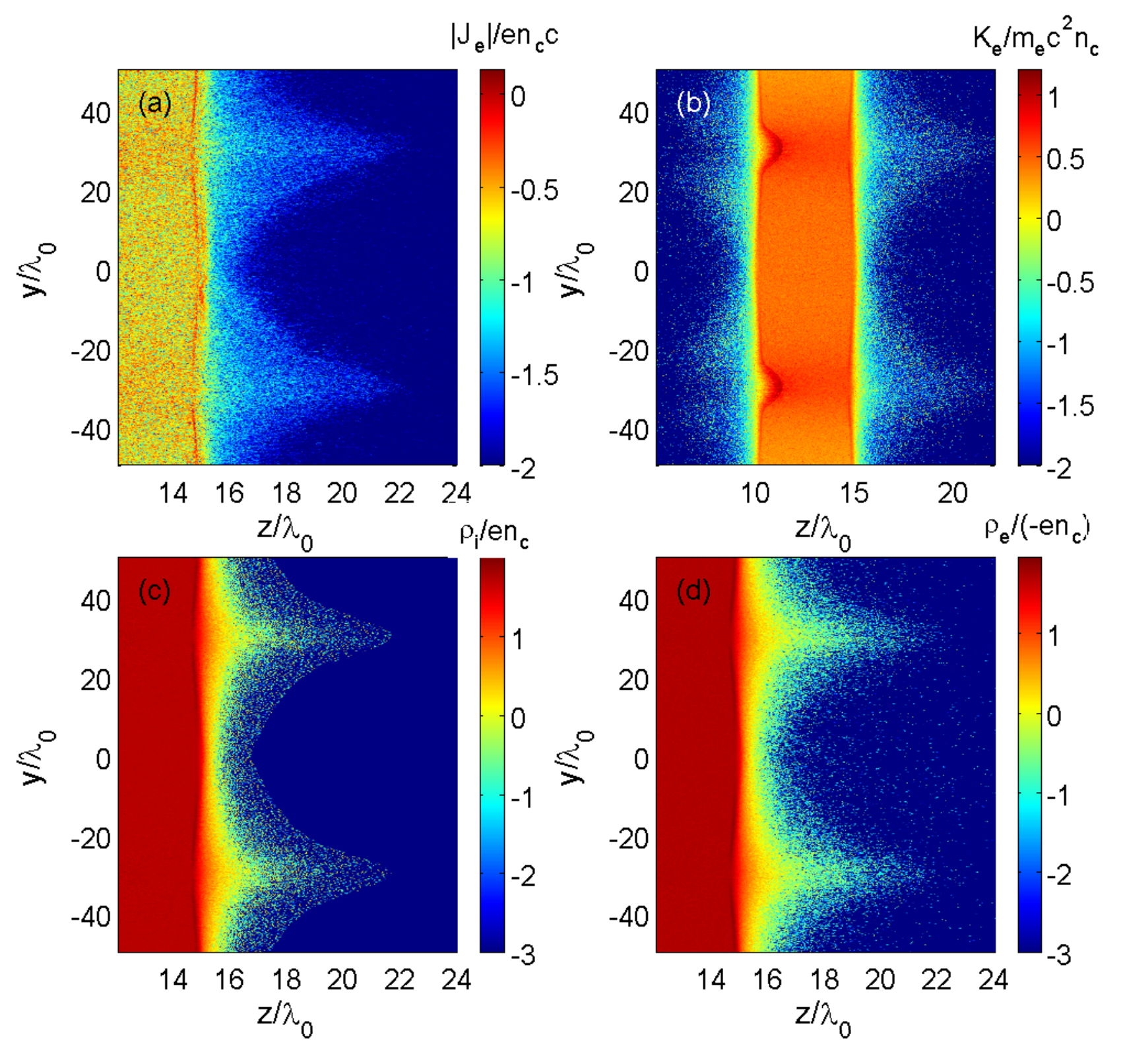}%
\caption{\label{f2} (Color online) The amplitude of the electron current
density ($|J_y+i*J_z|$) in the simulation plane (a), the electron kinetic
energy density in units of $m_ec^2n_c$ (b), the ion charge density in units
of $en_c$ (c) and the electron charge density in units of $-en_c$ (d).
The amplitude of the current density is in units of $en_cc$. All color
scales are 10-logarithmic and the simulation time is t=120$T_0$.}
\end{figure}

The impact of the double pulse generates two rarefaction waves, as can be
seen in Fig. \ref{f2}.
The electrons are accelerated mainly by $\mathbf{J}\times\mathbf{B}$ heating
(i.e., by the oscillating component of the laser's ponderomotive force)
\cite{s16} to a temperature of $\sim$1.0 MeV as the laser pulses irradiate the
target, which is close to Haines' scaling $T_h=m_ec^2(\sqrt{1+\sqrt{2}a_0}-1)=
1.01$MeV \cite{s17} with a velocity of $0.95c$. The electron temperature obtained
here is much lower than that given by ponderomotive scaling ($\sim$2.6 MeV)
\cite{Wilks92}. This discrepancy can be attributed to the fact that the electrons
only interact with the laser pulse during a fraction of the laser period before
being accelerated forward beyond the laser penetration region, due to that in our simulation that ultra-intense
laser pulses irradiate an overdense target with a steep density interface.
This situation is different from the mechanisms proposed by Kemp et al. \cite{Kemp09} and May et al.
\cite{May11}; a low density shelf is required
for the former and the electron energy is gained in the vacuum from the transverse
laser field in the latter case.

The hot electrons propagate through the target and form the two high energy density
channels visible in Fig. \ref{f2}(b) at $y=\pm 30\lambda_0$. A small part of the
energetic electrons escape into the vacuum at the rear surface of the target. Their
current, visualized in Fig. \ref{f2}(a), is not balanced by an ion current and generates
a strong sheath electromagnetic field, which reflects most of the hot electrons and
confines them. The sheath field accelerates ions on the target's surface (See Fig.
\ref{f2}(c)), whose larger inertia implies that they trail the electrons in Fig.
\ref{f2}(d). The plasma density of both parabolic expanding rarefaction waves decreases
with increasing distance from the target and the charge separation drives an ambipolar
electric field.

\begin{figure}\suppressfloats
\includegraphics[width=8.7cm]{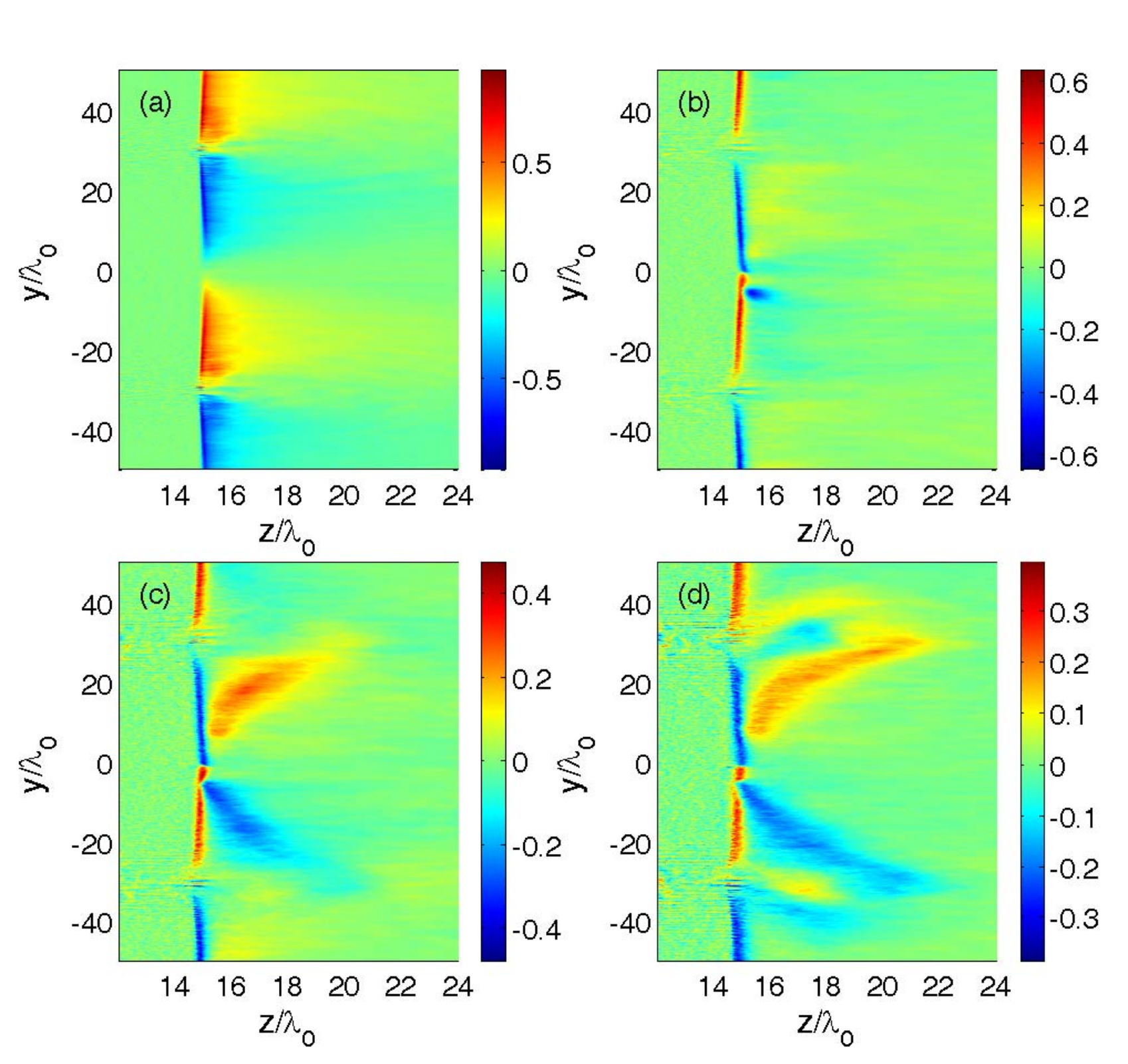}%
\caption{\label{f3} (Color online.) Transverse magnetic field ($B_x$) distribution at
t=60$T_0$ (a), 102$T_0$ (b), 120$T_0$ (c), and 140$T_0$ (d), respectively. The amplitude
is expressed in units of $m_ec\omega_0/e$.}
\end{figure}

The flow of hot electrons at the surface of the target and the return current, which flows
within the target to provide current closure, result in the growth of magnetic fields
orthogonal to the simulation plane, which is here the x-direction. In the case
of a single laser pulse, strong surface magnetic fields will expand from the exit point of
the electrons at the rear end of the target until they cover the surface uniformly. Here the
currents driven by both laser pulses will eventually collide and a more complex magnetic
topology is revealed by Fig. \ref{f3}.
The distribution of $B_x$ at the early time 60$T_0$ in Fig. \ref{f3}(a), which is triggered by
each laser pulse, equals that observed for single laser pulses. The hot electron's surface
current sheaths expand at the speed $0.7c$, but they have not yet collided at $y=0$. The
polarity of $B_x$ at $y= \pm 30\lambda_0$ switches, because the electrons above the
interaction point flow in the opposite direction compared with those below that point.

The hot electron sheaths have collided at the time 102$T_0$ in Fig. \ref{f3}(b). The surface magnetic
field has decreased in amplitude compared to that at 60$T_0$ and the field is more confined along z. A
sharp transition between the magnetic fields occurs at $y=0$ and a localized magnetic structure is
observed at $z\approx 15.5\lambda_0$ and $y\approx -4\lambda_0$. The polarity (negative amplitude)
suggests that it is tied to the electron sheath generated by the upper laser pulse. The current distribution
is not perfectly symmetric at this time as we may expect from our symmetric simulation setup. However,
the approximation of the plasma by computational particles introduces some randomness. Even a small
difference of the current or charge density of the surface electrons can result in significant differences
in the beam evolution, because their separation by the repulsive magnetic interaction corresponds to a
plasma instability. Statistical variations could explain the observed asymmetric current distribution.
Such differences do not limit the comparison of our PIC simulation results with that in experiments, since two identical
laser pulses would never be obtained in experiments.

This current structure and the magnetic field it drives have expanded further in Fig. \ref{f3}(c) ($t=120T_0$).
A second similar one has developed at $y>0$ and $z>15\lambda_0$. We call these structures magnetic wings.
The distribution of $B_x$ at the time 140$T_0$ in Fig. \ref{f3}(d) suggests that the currents responsible for
the wings are about to reconnect to the surface currents at $|y| <30\lambda_0$ and $z\approx 15 \lambda_0$.
We infer this from the magnetic field polarity. We find in Fig. \ref{f3}(d) a magnetic field patch with a
positive amplitude in the interval close to $y \approx 40\lambda_0$ and $z \approx 18\lambda_0$ and
in an interval close to $y \approx 30\lambda_0$ and $z \approx 20\lambda_0$. Both are about to merge.
Such a merger implies that the currents encircling both structures reconnect. An unambiguous demonstration
of such a reconnection can only be provided by the electronic currents. They are, however, too noisy for
this purpose. The magnetic field distribution is smoother, because it is connected to the well-resolved total
current and not to the fluctuating current densities within individual simulation grid cells.

The magnetic fields observed in Fig. \ref{f3} are transient. The current, which has been driven directly by
the laser pulses, will eventually be dissipated. The thermoelectric instability, an instability driven by a plasma
density gradient that is not parallel to the temperature gradient \cite{Tidman}, will eventually develop and
remagnetize the plasma. The dynamics of the magnetic fields it drives and their reconnection have been
examined in the experiments performed in Refs. \cite{s14a,s14b}. We do not consider them here.

\begin{figure}\suppressfloats
\includegraphics[width=8.7cm]{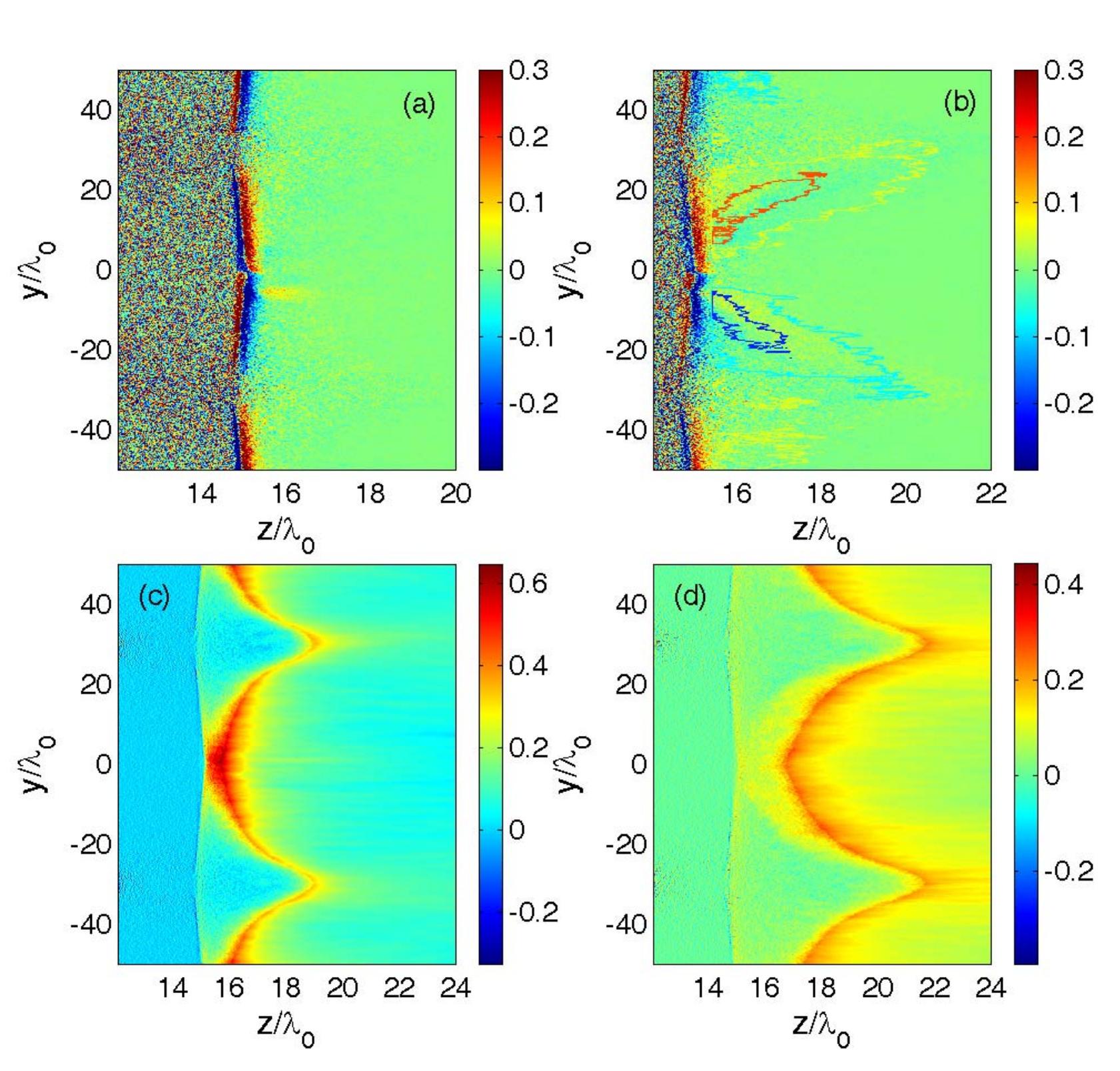}%
\caption{\label{f4} (Color online.) The transverse current density ($J_y$) distribution at t=102$T_0$ is
shown in (a) and that at 120$T_0$ in (b). Several contour lines of $B_x$ are overplotted in (b) using the
same color scale as for the current density. The electric field ($E_z$) distribution at t=90$T_0$ is shown in
(c) and that at 120$T_0$ in (d). The color scale of the current density is restricted within [-0.3 0.3] in order
to distinguish the current distribution clearly in the vacuum. The current density and the electric field amplitude
are expressed in units $en_cc$ and $m_ec\omega_0/e$, respectively.}
\end{figure}

Figure \ref{f4} displays $J_y$, which corresponds to the surface current along the initial target boundary,
and the electric field's $E_z$ component at two times (only the target's rear side is shown). The ambipolar
$E_z$ is tied to the density gradient of the rarefaction wave and it is here practically electrostatic.
Strong surface currents are present in Fig. \ref{f4}(a) at $z\approx 15\lambda_0$. The
fluctuations observed for $z< 15\lambda_0$ are thermal noise. No such fluctuations are seen for $z> 15
\lambda_0$, since the lower plasma density implies lower fluctuation amplitudes. The current filaments at
$z\approx 15\lambda_0$ have a thickness, which is a few times the electron skin depth $c/\omega_p \approx
0.023\lambda_0$. Consider the point $y=30\lambda_0$ in Fig. \ref{f4}(a). The surface current above this
point is negative, which implies that electrons move to increasing values of $y$. It has the opposite sign
below. The hot surface electrons thus move away as expected from the point $y = 30\lambda_0$. In addition,
a lower return current density at $z <15\lambda_0$ is visible. Both currents encircle the surface magnetic field
visible in Fig. \ref{f3}(b).

The surface current in Fig. \ref{f4}(a) maintains a practically constant strength from $|y|=30\lambda_0$ up to
$y=0$. The electron flow has not been thermalized by an electrostatic instability, as this would demagnetize the
flow around $y=0$ and increase the electron temperature. The currents have been separated instead along the
normal direction of the target's rear surface in a small y-interval close to $y\approx 0$. The surface current driven
by the upper laser pulse is magnetically expelled at $y \approx -4\lambda_0$ and $z\approx 15.5\lambda_0$ and
$J_y$ reaches here a value $5.4\times10^{11}$A/cm$^2$. The total current density is actually stronger, because
the expelled current has $J_y$ and $J_z$ components due to its deflection along z. Its spatial correlation with the
magnetic structure in Fig. \ref{f3}(b) indicates that the expelled current drives it. The surface current driven by
the lower laser pulse initially reconnects with the return current driven by the upper laser pulse. We attribute
again the different behaviour of the electron flows driven by the upper and lower laser pulses to the break of
symmetry by the plasma approximation by a finite number of computational particles and the resulting statistical
variations. The symmetric wings in Fig. \ref{f3}(c) evidence that both current sheaths get expelled at a later time.
Note that we can not clearly associate a current with the overplotted contour lines of the wings in Fig. \ref{f4}(b),
because the current is too weak and noisy. The total current is, however, preserved since Fig. \ref{f3}(c) clearly
demonstrates that the peak amplitudes of the surface current and wing magnetic field are practically identical.

The reason for the widening of the expelled current sheath, which is responsible for the spreading out of the
magnetic wing with increasing distance from the target surface in Fig. \ref{f3}(c), is given by Figs. \ref{f4}(c) and
\ref{f4}(d). The parabolic ambipolar electric field visible in both plots outlines the front of the two rarefaction waves,
which are driven by the laser pulses. This interpretation is supported by the electron and ion distributions in Figs.
\ref{f2}(c) and \ref{f2}(d). Note that no ambipolar electric field can be observed at the original target boundary
$z = 15\lambda_0$, because the strong surface magnetic field suppresses the electron mobility along the surface
normal. A comparison between Fig. \ref{f3}(c) and Fig. \ref{f4}(d) demonstrates that the magnetic wings are located
within the rarefaction wave. A current closure is thus achieved by the rarefaction wave's plasma. The plasma density
within the rarefaction wave is below that of the solid target, which reduces the supported current density. The currents
and the associated magnetic fields spread out in space.

\begin{figure}\suppressfloats
\includegraphics[width=8.7cm]{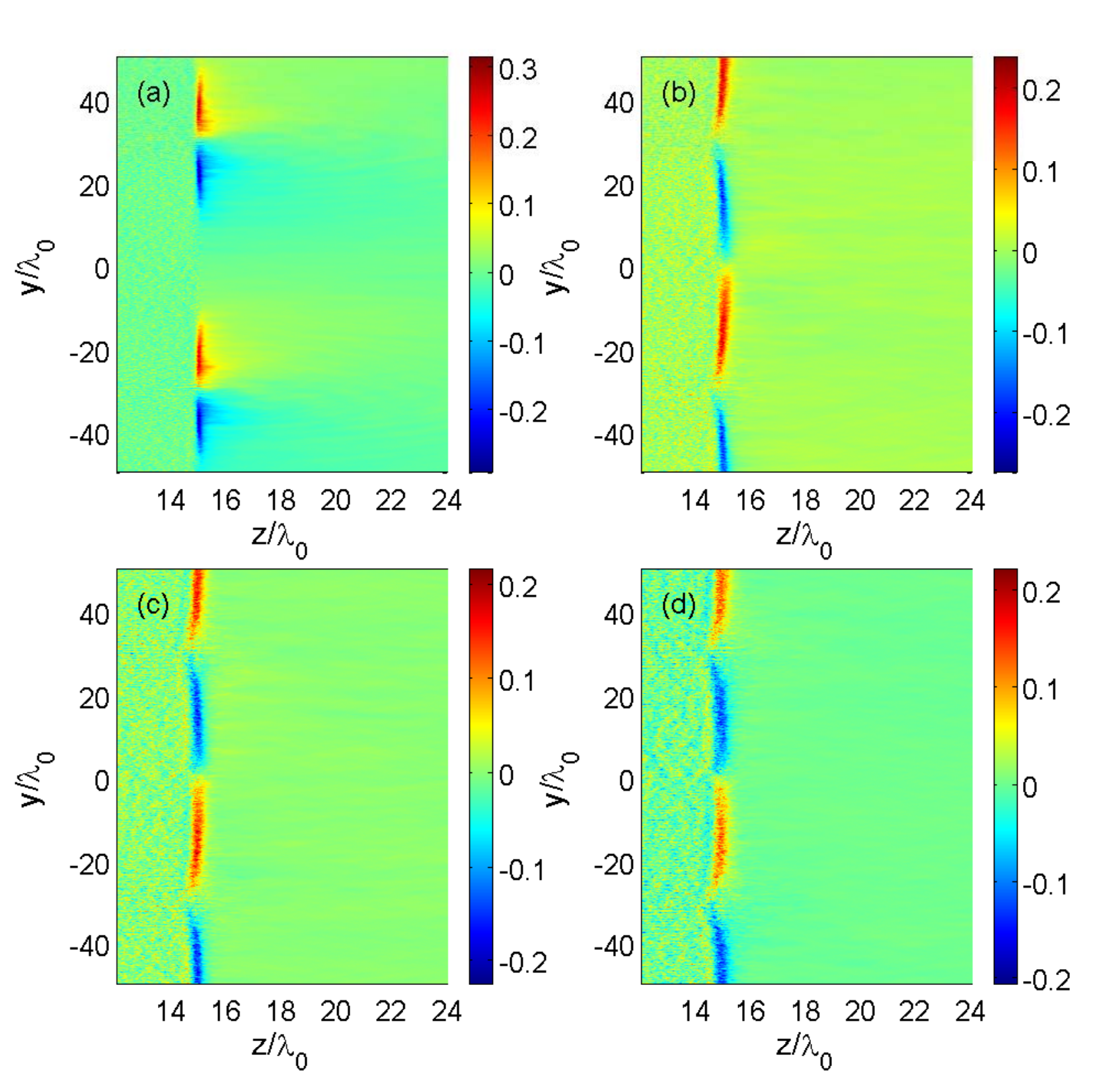}%
\caption{\label{f5} (Color online.) Transverse magnetic field ($B_x$) distribution at t=60$T_0$ (a), 100$T_0$ (b),
120$T_0$ (c), and 140$T_0$ (d), respectively, for the case with a laser intensity of $5\times10^{18}$W/cm$^2$.
The other parameters are the same as that in Fig. \ref{f3}}
\end{figure}

It is interesting to know how the laser power and, thus, the characteristic speed of the electron sheath affect the
interaction of the electrons and the magnetic fields in the sheaths' overlap layer. More specifically, we want to know
if the filamentation instability could also be examined with weaker laser pulse intensities, like those used in the previous
experiments \cite{s14a,s14b}. We have performed for this purpose a series of simulations with lower laser intensities,
i.e., $I_0=5\times10^{16}$W/cm$^2$, $5\times10^{17}$W/cm$^2$, and $5\times10^{18}$W/cm$^2$, respectively.
The magnetic field repulsion in the sheath overlap layer is not observed in any of the
simulations. Figure \ref{f5} shows the evolution of the transverse magnetic field for the case study that employed a
laser intensity $5\times10^{18}$W/cm$^2$.
The magnetic field amplitude driven by the surface current is much weaker than that observed for a laser intensity $5\times10^{19}$W/cm$^2$ (Fig. \ref{f4}). The maximum magnetic field here is $\sim2.8\times10^{7}$Gauss at $t=100T_0$, only one third of the latter ($\sim7.4\times10^{7}$Gauss). Both the current density and the expansion
speed of the electron current along the surface here ($\sim0.4c$) are much smaller than that of the latter. Most
importantly, Fig. \ref{f5} reveals a different magnetic topology close to $y\approx 0$ compared to Fig. \ref{f3}.
The electrons from both sheaths do not interpenetrate in the simulation that uses the weaker laser pulse. This implies
that the driving currents are closed by the return current within the target. A magnetic distribution develops at the
surface, which is stationary during the resolved time interval. The absence of counter-streaming electron beams implies
that no filamentation instability can develop here. If the electron beams would interpenetrate close to $y\approx 0$,
their nonrelativistic speeds $\approx 0.4c$ observed in the simulation would probably result in a thermalization
through the quasi-electrostatic two-stream and oblique modes \cite{Filament}.

\section{CONCLUSION}
In summary, we have examined the plasma processes triggered by the impact of a laser double-pulse on a solid
target, using PIC simulations. The large 2D simulation box has resolved a cross section of the target. Each laser pulse
has accelerated the electrons to relativistic speeds. The electrons have mainly been accelerated along the propagation
direction of the laser pulse and they have crossed the target. The electrons re-emerged on the target's rear surface,
where they have been deflected by the self-generated ambipolar electrostatic field. They have expanded along the
target's surface and they have collided at the midpoint between the laser axes. The plasma did not thermalize electrostatically when the electron sheaths collided. Their relativistic speeds resulted in a dominant magnetic rather
than electrostatic interaction. Our 2D geometry implied that the current sheaths were expelled from the target by
their mutual interaction and moved into the rarefaction wave, provided that the driving laser pulse is ultra-intense.
The resolution of a second dimension of the target's surface by a currently too expensive 3D PIC simulation would allow
for studies of the filamentation instability \cite{New1,New2,New3} within the overlap layer of the colliding electron
sheaths. Our simulation results thus suggest that this instability, which is thought to magnetize energetic astrophysical
flows, can be observed in a laboratory experiment provided that the laser pulse is ultra-intense and the electron flow
speeds are relativistic.

\begin{acknowledgments}\suppressfloats
This work was supported by EPSRC (Grant No. EP/D06337X/1) and partly
supported in the framework of the HiPER consortium. M.B. also acknowledges funding from projects ELI (Grant No. CZ.1.05/1.1.00/483/02.0061) and
OPVK 3 (Grant No. CZ.1.07/2.3.00/20.0279). X.H.Y. acknowledges the support from
the China Scholarship Council, the NSFC (Grant Nos. 10975185 and 10976031), the Innovation
Foundation for Postgraduate of Hunan Province (Grant No. CX2010B008) and NUDT (Grant No. B100204).
G.S. wishes to acknowledge the Leverhulme Trust (Grant No. ECF-2011-383).

\end{acknowledgments}

\end{document}